\documentclass[10pt,nofootinbib,showpacs,twocolumn,epsfig,aps,prd,amsmath,amssymb,superscriptaddress,showpacs]
{revtex4-1}
\usepackage[utf8]{inputenc}

\usepackage{epsfig}
\usepackage{chngcntr}
\usepackage{wrapfig}

\usepackage{dcolumn} 
\usepackage{bm}            
\usepackage{subfigure}

\usepackage{longtable}
\usepackage{amsbsy,amsmath}
\usepackage{amssymb}
\usepackage{hyperref}
\usepackage{slashed}

\usepackage{graphicx}

\newcommand{\beq}{\begin{eqnarray}}

\newcommand{\eeq}{\end{eqnarray}}

\newcommand{\be}{\begin{equation}}

\newcommand{\ee}{\end{equation}}

\def\fun#1#2{\lower3.6pt\vbox{\baselineskip0pt\lineskip.9pt
\ialign{$\mathsurround=0pt#1\hfil ##\hfil$\crcr#2\crcr\sim\crcr}}}

\newcommand{{\SD}}{\rm SD}

\newcommand{{\Mc}}{\mathcal{M}}

\newcommand{\ver}{\mbox{\boldmath${\rm r}$}}

\newcommand{\vep}{\mbox{\boldmath${\rm p}$}}
\newcommand{\veq}{\mbox{\boldmath${\rm q}$}}

\newcommand{\vek}{\mbox{\boldmath${\rm k}$}}

\newcommand{\vegam}{\mbox{\boldmath${\rm \gamma}$}}

\renewcommand\Im{\operatorname{Im}}

\begin{document}

\title{Fluctuation sound absorption in quark matter }


\author{B.O. Kerbikov}
\email{borisk@itep.ru}

\affiliation{{\sc Alikhanov Institute for Theoretical and Experimental Physics},\\ Moscow 117218, Russia \medskip}

\affiliation{{\sc Moscow Institute of Physics and Technology},\\ Dolgoprudny 141700, Moscow Region, Russia \medskip}

\affiliation{{\sc Lebedev Physical Institute},\\ Moscow 119991, Russia \medskip}

\author{M.S. Lukashov \medskip}
\email{lukashov@phystech.edu}

\affiliation{{\sc Alikhanov Institute for Theoretical and Experimental Physics},\\ Moscow 117218, Russia \medskip}

\affiliation{{\sc Moscow Institute of Physics and Technology},\\ Dolgoprudny 141700, Moscow Region, Russia \medskip}

\date{\today}

\begin{abstract}

We investigate the sound absorption in quark matter due to the interaction of the sound wave with the precritical fluctuations of the diquark-pair field above $T_c$. The soft collective mode of the pair field is derived using the time dependent Ginzburg-Landau functional with random Langevin forces. The strong absorption near the phase transition line may be viewed as a manifestation of the Mandelshtam-Leontovich slow relaxation time theory.   

\end{abstract}

\pacs{21.65.Qr, 12.38.Mh, 25.75.Nq}

\maketitle


\section{Introduction}
QCD under extreme conditions has  been a subject of intense study for the
last decade. While the properties of quark-gluon matter at high temperature
and zero   chemical potential are theoretically investigated in great
detail, understanding of the quark matter physics in the regime of non-zero
density and moderate or low temperature remains challenging. This is due to
the fact that the high $T$ and zero $\mu$ region of the QCD phase diagram is
accessible to lattice simulations. In the non-zero density regime
Monte Carlo simulations fail and one has to  resort to  models, like the NJL
one. On the  experimental side,  information obtained at RHIC and LHC
corresponds in bulk to the high temperature and low density region. Non-zero
density and moderate or low temperature conditions may exist  in neutron
stars and will be possibly realized in future experiments at FAIR and NICA.

According to the present understanding of the QCD phase structure, 
attractive interaction between quarks in color antitriplet state
leads to the  formation of the color superconducting phase in the moderate
and high density domain \cite{1,2,3}. Some important features of this phase
are, however, very  different from that of the BCS superconductor \cite{4}. In
particular, instead of an  almost sharp dividing line between the normal and
superconducting phases in the BCS case, in color superconductor the
transition is significantly smeared. Correspondingly, an exceedingly  narrow
precritical fluctuation region in the BCS superconductor is replaced by a
rather wide and physically important one in color superconductor. The
fluctuation contribution to the physical quantities is characterized by the
Ginzburg-Levanyuk number  $Gi$ which for the quark matter may be  estimated
as \cite{4}

\be Gi\simeq\frac{\delta T}{T_c} \simeq \left( \frac{T_c}{\mu}\right)^4
\simeq10^{-4},\label{1}\ee
where $T_c \cong 40$ MeV in the  critical temperature for  the $2SC$ phase
($u$ and $d$ quarks pairing), $\mu\cong 400$ MeV  is the quark chemical
potential. Note that for the BCS superconductor $Gi\sim 10^{-12} -
10^{-14}$. Fluctuation quark pairing which takes place when the temperature
approaches $T_c$ from above manifests  itself in the characteristic
temperature   dependence of a number of physical quantities. In BCS
superconductors such phenomena have been intensively studied for more than
three decades \cite{6}. In  our previous paper \cite{5} we have calculated
the fluctuation electrical  conductivity of quark matter. It has been shown
it is large and greatly exceeds the Drude one.

In the present paper we investigate the fluctuation sound absorption and
show that it has  even more pronounced temperature dependence than the
electrical  conductivity. Let us point from the very beginning that we
consider the hydrodynamical, or first, sound.

The reason for the strong  energy dissipation of the sound wave  in the
precritical region has  a general nature. The idea goes back to the seminal
paper \cite{7} in which Mandelshtam and Leontovich formulated the slow
relaxation time theory (see also \cite{8,9,10,11,12}). Suppose that the
relaxation time corresponding to the equilibrium restoration is  large. Then
during the equilibration process strong energy dissipation occurs.
Propagation of the sound wave changes the  critical temperature in the
compression -- rarefraction regions. The fluctuation pairing is a slow
process and the resulting nonequilibrium results in the sound wave energy
loss \cite{13,14}. Based on the  above ideas we shall calculate the
fluctuation sound absorption in quark matter. Unlike other transport
coefficients (shear  viscosity, electrical conductivity, etc.),  sound
propagation in quark matter did not  receive much  attention in the
literature. The quantity ($1/3 - c^2_s$) reflects the breaking of the
conformal symmetry and may serve as  a measure of the interaction. Lattice calculation of this
quantity may be found  in \cite{15}. The squared speed of   the  sound $c^2_s$
as a function of $T$ at zero density was also calculated in this  work and
the minimum which corresponds to the softest point of the EoS  was found.
Different problems related to sound propagation in quark-gluon matter have
been discussed in \cite{16}.

The paper is structured as follows. In Sec.II we introduce the time dependent Ginzdurg-Landau functional with random Langevin forces and derive the fluctuation propagator (FP) In Sec.III we present the Aslamazov-Larkin (AL) diagram for the polarization operator. Using the expression for the FP derived in Sec.II we evaluate the AL diagram and show that it gives rise to a strong sound absorption near the critical temperature. The final Sec.IV is devoted to a summary and concluding remarks. \smallskip

\section{Collective Mode Propagator}

Throughout this paper we use the natural system of units $\hbar=c=k_B=1$. The
Matsubara fermion propagator has the form \be G(\vep, \varepsilon_n) =
\frac{1}{\gamma_0 (i \varepsilon_n +\mu) - \vegam \vep -m}.\label{2}\ee Here
$\varepsilon_n =\pi T (2n +1), \mu$ is the quark chemical potential.
Integration in the vicinity of the Fermi surface is performed making use of the
variable $\xi $ defined as \be \xi = \sqrt{\vep^2+m^2} -\mu.\label{3}\ee Then
\beq
\nonumber\int \frac{d\vep}{(2\pi)^3} &\simeq&  \int  d\xi\rho (\xi) \simeq \\ 
\nonumber &\simeq& \int d\xi\left[\rho (\mu) + \left( \frac{\partial\rho}{\partial\xi}\right)_\mu \xi\right]d\xi=\\
&=&\frac{p_0\mu}{2\pi^2} \int d\xi + \frac{\mu}{2\pi^2}\left(\frac{v^2_0+1}{v_0}\right) \int d\xi\,\xi. 
\label{4}
\eeq
Here $p_0$ is the Fermi momentum, $v_0=p_0/\mu$ is the Fermi velocity. The
second term in (\ref{4}) takes into account the energy dependence of the density
of states at the Fermi surface. As will be shown below only the contribution
from this  term enters into the final result for the fluctuation sound
absorption.

Fluctuations of the pair field in the  vicinity of $T_c$ are described by the collective mode, or the fluctuation propagator (FP) \cite{6}.  In \cite{6} and references therein the FP $L(\veq, \omega_k)$ was introduced in the framework of the BCS theory making
use of the nonrelativistic kinematics and Green's functions. In \cite{5} it was
derived for the relativistic quark  system solving the Dyson equation with
Matsubara propagators (\ref{2}). Graphically the Dyson equation is represented
in Fig.1.

\begin{figure}[h]\centering
\begin{center}
\resizebox{0.95\columnwidth}{!}{\includegraphics[scale=1.0]{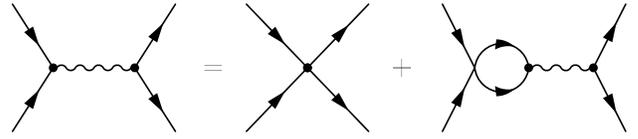}}
\end{center}
\vspace{-3mm}
\caption{Dyson equation for FP (wavy line).}
\label{fig:FIG1}
\end{figure}

Here the FP will be obtained using the time-dependent Ginzburg-Landau
functional TDGL \cite{17, 18, 6} and the  stochastic Langevin forces. In absence of  external electromagnetic field the TDGL for the fluctuating pair
field $\psi$ reads \be -\gamma \frac{\partial}{\partial t} \psi (\ver, t) =
\frac{\delta F[\psi]}{\delta \psi*} +\eta (\ver, t). \label{5}\ee

Here $\gamma$ is the order parameter relaxation time constant. The GL
functional with the quartic term dropped has the form \be F[\psi] =F_0 + \int
d\ver\left\{ a | \psi (\ver, t ) |^2 + b |\nabla\psi (\ver, t)
|^2\right\},\label{6}\ee where $a =\nu\varepsilon,\,\, \nu=\rho (\mu) =
\frac{p_0\mu}{2\pi^2}, \varepsilon=\frac{T-T_c}{T_c}, b=\nu\varkappa^2, \varkappa^2 =
\frac{\pi}{8 T_c} D, D$ is the diffusion coefficient, $\gamma= \frac{\pi\nu}{8
T_c}$ (for details see \cite{17,18,6,4} ). With $F[\psi]$ given by (\ref{6}) we
return to (\ref{5}) and write 

{\small
\be {-\left[ \gamma\dfrac{\partial}{\partial t} + \nu\bigl( \varepsilon + \varkappa^2\veq^2 \bigr) \right]\Psi(\ver, t) \equiv {\hat L}^{-1} \Psi(\ver,t) = \eta(\ver, t).}\label{7}\ee
}

The solution of (\ref{7}) may be formally written as

\be \Psi(\ver, t) = {\hat L}\eta(\ver, t).\label{8}\ee 

Let us assume that the correlator of the Langevin forces has a gaussian form

\be \left< \eta^*(\ver, t)\eta(\ver', t') \right> = \gamma\,\delta(\ver-\ver')\,\delta(t-t'). \label{9}\ee

According to the fluctuation-dissipation theorem \cite{18} the retarded propagator (the FP in our case) is given by the equal time correlator $\left< \Psi^*(\ver, t)\Psi(\ver', t) \right>$. From (\ref{8}) one can write 

\be \begin{split}\left< \Psi^*(\ver, t)\Psi(\ver', t) \right>\,&=\,\gamma\int\dfrac{d\veq}{(2\pi)^3}\,e^{i \veq (\ver-\ver')}\times \\ \times \int\limits_{-\infty}^{+\infty}\dfrac{d\omega}{2\pi} {\hat L}^{*}&(\ver, \omega){\hat L}(\ver, \omega)= \\
= -\int\dfrac{d\veq}{(2\pi)^3}\,e^{i \veq (\ver-\ver')}&\int\limits_{-\infty}^{+\infty}\dfrac{d\omega}{2\pi}{\omega}^{-1}\Im{\hat L}(\ver, \omega),\label{10}\end{split}\ee where \be {\hat L}(\ver, \omega)=-\dfrac{1}{\nu}\,\dfrac{1}{\varepsilon + \frac{\pi}{8 T_c}\bigl(-i\omega+D\veq^2\bigr)}.\label{11} \ee

The FP (\ref{11}) describes the slow diffusion mode near the critical
temperature. At small $\omega$ and $q$ close to $T_c$ the  quantity  $L(\veq,
\omega)$ can be arbitrary large and is rapidly varying. This will be an
important point in the calculation of the Aslamazov-Larkin diagram for the
sound absorption.\smallskip

\section{Precritical sound absorption}
In this work we study the effects caused by the quark pair field fluctuations.
The possible role of  the gluon field fluctuations has been studied in detail
in \cite{4} and also in \cite{20}. According to \cite{4} the gluon field
fluctuations lead to a shift in $T_c$ and to possible replacement of the
second-order phase transition to the first-order one. However, the increase of
the quark density leads to a suppression  of the gluon fluctuations \cite{4}.
The authors of Ref. \cite{20} also came to the conclusion that the fluctuations
of the pair field dominate those of the gauge field in the strong coupling
regime.

With the FP at our disposal, we can evaluate the Aslamazov-Larkin (AL)
\cite{13,14} contribution to the sound absorption  in  the fluctuation region.
Based on  the experience  gained in condensed matter physics \cite{6}, we
assume  that it is of major importance among other  quantum  fluctuation
effects. Previously it was shown that AL paraconductivity exceeds the Drude one
\cite{5}. In \cite{21} preliminary  results on the AL term in lepton-pair production
were presented.

The Feynman diagram representing the AL sound absorption is shown in Fig.2. 
\begin{figure}[h]\centering
\begin{center}
\resizebox{0.70\columnwidth}{!}{\includegraphics[scale=1.0]{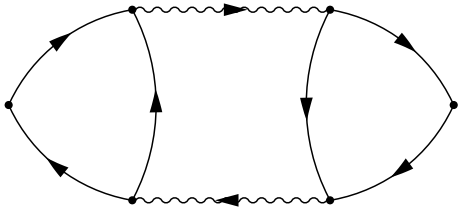}}
\end{center}
\vspace{-5 mm}
\caption{\label{fig:FIG2}Feynman diagram for the AL polarization operator for the sound absorption.}
\end{figure}
It contains two wavy  lines corresponding to the FP and this makes this contribution  the most
important in the vicinity of $T_c$. The sound absorption is determined by the
imaginary part of the polarization operator given by the AL diagram. The in- and  out-  vertices in this diagram are equal to the  constant
$g$ of the phonon-quark interaction. The solid lines are the quark  propagators
(\ref{2}). The AL diagram corresponds to the following polarization operator

\be\begin{split} {\Pi} (\vek, \omega_\nu) = 4 T \sum_{\Omega_j} \int &\frac{d\veq}{(2\pi)^3}
B^2 (\vek,\veq, \omega_\nu , \Omega_j)\times\\ \times L(\veq+\vek, &\Omega_j + \omega_\nu)
L(\veq, \Omega_j).\label{12}\end{split}\ee

Here $\vek$ is the sound wave momentum, $\omega_\nu$ is the Matsubara sound
frequency, $B(\vek, \veq, \omega_\nu, \Omega_j)$ is the block of the  three
propagators

\be\begin{split}B(\vek, \veq, \omega_\nu, \Omega_j) = g\,T\,\sum_{\varepsilon_n} &\int
\frac{d\vep}{(2\pi)^3}\times\\ \times \text{tr} \bigl\{ G (\vep, \varepsilon)G(\vep+ \vek,
\varepsilon_n +\omega_\nu)&G(\veq -\vep,\Omega_j
-\varepsilon_n)\bigr\}.\label{13}\end{split}\ee

The dependence of the FP-s  $L(\vek+\veq, \Omega_j + \omega_\nu)$ and $L(\veq,
\Omega_j)$ on $\Omega_j $ and $\omega_\nu$ is much stronger than that
of the Green's functions. The closeness to the transition point is enclosed in the FP-s at small values of frequencies and momenta.  Therefore we shall keep in the
propagators entering into $B$ only the dependence on the fermionic frequencies
and momenta. In this approximation one easily obtains

\be\begin{split} \text{tr} \bigl\{ G(\vep, \varepsilon_n)&\,G(\vep, \varepsilon_n)\,G(-\vep, - \varepsilon_n)\bigr\} =\\ 
\frac{2m}{E}&\frac{1}{(\xi-i\varepsilon_n)^2(\xi+ i \varepsilon_n)}
\simeq\\ \frac{2 m}{\mu}&\frac{1}{(\xi-i\varepsilon_n)^2(\xi+
i \varepsilon_n)}.\label{14}\end{split}\ee  
Next we transform according to (\ref{4}) the
integration over $d\vep$ in (\ref{14}) into the integration over $d\xi$ and
perform the summation over $\varepsilon_n$. Only the second term in (\ref{4})
proportional to $\left( \frac{d\rho}{d\xi}\right)_\mu$ gives nonzero
contribution in the integral over $d\xi$. The result for $B$ yields

{\small
\be\begin{split} B = g\,T\,\frac{2m}{\mu}&\sum_{\varepsilon_n}\int d\xi
\frac{\xi}{(\xi-i\varepsilon_n)^2(\xi+ i \varepsilon_n)}\left(\frac{\partial\rho}{\partial\xi}\right)_\mu = \\&g\ \frac{m}{2\pi^2}\left(\frac{v_0^2+1}{v_0} \right) ln \frac{\omega_D}{2\pi T_c}.\label{15}\end{split}\ee}

The critical temperature for 2SC superconducting phase is $T_c\simeq 40 $
Mev, the  ultraviolet cutoff $\omega_D\simeq 800$ MeV \cite{2,3,4}, therefore
$ln \frac{\omega_D}{2\pi T_c}\simeq 1.$

Upon the substitution of (\ref{15}) into (\ref{12}) one gets
{\small
\be\begin{split} \Pi (\vek, \omega_\nu) &= g^2\frac{m^2}{\pi^4}
\left(\frac{v_0^2+1}{v_0} \right)^2\,\ln^2\,\frac{\omega_D}{2 \pi T_c}\times\\ \times T\,\sum_{\Omega_j} \int &\frac{d\veq}{(2\pi)^3} L(\veq+\vek, \Omega_j + \omega_\nu)
L(\veq, \Omega_j).\label{16}\end{split}\ee}

To proceed further, we shall assume that the acoustic wavelength is much larger than the correlation radius of fluctuations, i.e.,

\be \varepsilon \gg \dfrac{\pi}{8 T_c} D k^2. \label{17}\ee

For $T\tau \ll 1$ the diffusion coefficient is $D=\frac{1}{3}v_0^2\tau$. As an order of magnitude estimate we take $\tau \simeq 0.3 fm$, $T_c=40 MeV$, $v_0^2=\frac{1}{3}$. Then 

\be \dfrac{k^2}{MeV^2} \ll 10^6 \varepsilon. \label{18}\ee

Therefore (\ref{17}) imposes a very weak restriction on the phonon momentum. The inequality (\ref{17}) allows to neglect the $\vek$-dependence of the $FP$ $L(\veq+\vek, \Omega_j + \omega_\nu)$. To evaluate the sum over $\Omega_j$ in (\ref{16}), we can use a technique of replacing the summation by contour integration \cite{22,23}. At the first step, this leads to the following result for the polarization operator 

\be\begin{split} \Pi (\omega) = \dfrac{2 B^2}{\pi}\int &\dfrac{d\veq}{(2\pi)^3}\int\limits_{-\infty}^{+\infty}\,dz\,\text{coth}\,\frac{z}{2T}\Bigl[L^R(\veq, -iz-i\omega)+ \\  + L^A& (\veq, -iz+i\omega) \Bigr] \Im L^R (\veq, -iz), \label{19}\end{split}\ee where $z=i \Omega_j$, $\omega=i \omega_{\nu}$, and $L^R$ and $L^A$ are the retarded and advanced $FP$-s. The next step is to expand the integral in powers of $\omega$ and to substract the zeroth order term which would lead to Meissner effect above $T_c$. Alternatively, this may be regarded as imposing the Word identity on the polarization operator. Keeping in (\ref{19}) the term proportional to $\omega$ and integrating by parts, one has 

\be\begin{split} \Pi (\omega) = -i \omega B^2 &\frac{4T}{\pi}\int\dfrac{d\veq}{(2\pi)^3}\int\limits_{-\infty}^{+\infty} dz \frac{(\Im L^R)^2}{z^2}=\\ -i\omega \dfrac{\pi B^2}{\nu^2}&\int\dfrac{d\veq}{(2\pi)^3}\dfrac{1}{\bigl( \varepsilon + \frac{\pi}{8 T_c} D \veq^2 \bigr)^3}. \label{20}\end{split}\ee

The final result for the fluctuation sound absorption coefficient reads 

\be\begin{split} \Im \Pi =& -\omega\,g^2\,\dfrac{m^2}{2^5 p_0^4 \varkappa^3}\,(v_0^2 + 1)^2\times \\ \times &\ln^2 \frac{\omega_D}{2 \pi T_c} \left(\dfrac{T_c}{T-T_c}\right)^{3/2}, \label{21}\end{split}\ee

where $\varkappa^2 = \dfrac{\pi}{8\,T_c} D$ (see definition following Eq. (\ref{6}))\smallskip

\section{Concluding Remarks}

In the paper we have examined the sound absorption in quark matter at moderate density and temperature due to the precursory fluctuations of the the quark pair field. The absorption is caused by the interaction of phonons with the soft collective mode of the quark field. The results are in line with the Mandelshtam-Leontovich slow relaxation time theory. The sound propagation changes the critical temperature in the compression-rarefaction regions. The fluctuation pairing is a slow process and the resulting inequilibrium leads to the intense sound wave energy loss. Assumptions which have been made in the course of the derivation were clearly exposed. The dependence of the sound absorption on the proximity to the critical temperature is even more pronounced than in case of the electrical  conductivity \cite{5}, where it was possible to compare the fluctuation contribution with the Drude one.  For the quark matter we are not aware of the ``normal'' sound absorption calculations. For the BCS superconductor it was shown that the fluctuation sound absorption exceeds the normal one in a wide range of parameter \cite{13, 14}.

The authors are supported by a grant from the Russian Science Foundation project number 16-12-10414. The authors express their deep gratitude to A.~Varlamov for illuminating discussians. We thank M.~Andreichikov, Ya.~A.~Simonov, L.~McLerran and P.~Petreczky for valuable remarks. \bigskip



\end{document}